# Topological Constraint Theory and Rigidity of Glasses

Mathieu Bauchy[1],*

[1]*Physics of AmoRphous and Inorganic Solids Laboratory (PARISlab)*
*University of California, Los Angeles, California, 90095, USA*
*bauchy@ucla.edu

**Abstract**

Topological constraint theory has become an increasingly popular tool to predict the compositional dependence of glass properties or pinpoint promising compositions with tailored functionalities. This approach reduces complex disordered networks into simpler mechanical trusses. Thereby, topological constraint theory captures the important atomic topology that controls macroscopic properties while filtering out less relevant second-order structural details. As such, topological constraint theory can be used to decode the genome of glass, that is, to identify and decipher how the basic structural building blocks of glasses control their engineering properties—in the same way as the human genome offers information that serves as a blueprint for an individual's growth and development. Thanks to its elegance and simplicity, topological constraint theory has enabled the development of various physics-based models that can analytically predict various properties of glass. In this Chapter, I introduce some general background in glass science, concepts of atomic rigidity, and topological constraint theory. The topological constraints enumeration scheme is presented for various archetypical glasses and is used to understand the origin of their glass-forming ability. Finally, various topological models enabling the prediction of glass properties are reviewed, with a focus on hardness, fracture toughness, viscosity, fragility, glass transition temperature, and dissolution kinetics.

1. **Introduction**
*1.1 The glass age*

As illustrated recently by the Materials Genome Initiative, the discovery of new materials with tailored functionalities is essential to economic development and human well-being [1]. In fact, 10 of the 14 societal Grand Challenges identified by the National Academy of Engineering are expected to require the development of novel materials with improved properties [2]. To this end, the discovery of new materials or the optimization of existing ones requires a deep understanding of how the composition and structure of materials control their macroscale properties.

The discovery of new materials has always played a crucial role in human history—to the point that human history time periods are named after materials: Stone Age, Bronze Age, and Iron Age (see **Fig. 1**). Among the many materials that have been discovered, glass has been one of the most influential and its importance keeps increasing [3]. Since the Romans started to use glass to make building windows, glass has defined human progress in many ways [4]. Spectacles have allowed individuals to recover their vision. Glass mirrors largely contributed to defining the concept of individual identity. Telescope lenses enabled major discovery in astronomy. Bacteria were discovered thanks to magnifying glasses. Glass was key in the development of light bulbs and television. Today, glasses are used to immobilize nuclear waste or stimulate bone growth after fracture [5]. Touch-screen display glasses have changed the way humans interact with electronic devices and virtually everybody carries a piece of glass in his pocket [6]. Maybe even more importantly, glass optical fibers made it possible for the development of the Internet as we know it today [7]. Glass also shows great promises to solve some of tomorrow's grand challenges in clean energy, environment, water treatment, healthcare, information, transportation, etc. [6]. For all these reasons, it has recently been proposed that we are now living in the ***Glass Age*** [8].

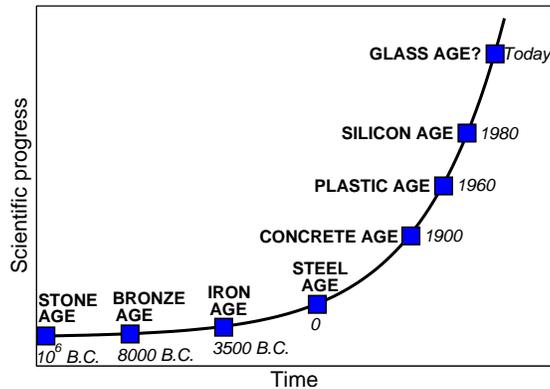

**Fig. 1:** The discovery of new materials largely defines the progress of our civilization.

### *1.2 Glass genome and discovery of new compositions*

Revealing the full potential of glass requires the discovery of new glass formulations showing unique properties. However, this task is especially complicated for non-crystalline glassy materials for several reasons. First, virtually all the elements of the periodic table can be turned into a glass, if cooled fast enough from the liquid state [9]. Second, unlike crystals, glasses do not have to satisfy any fixed stoichiometry thanks to their out-of-equilibrium nature. As such, the composition of glasses can be continuously changed. For all these reasons, the number of possible glass compositions has been estimated to be around $10_{52}$! Yet, only about 200,000 glass compositions have been produced in the last 6000 years of human glass history [9]. These numbers demonstrate that the range of possible glasses remains largely unexplored—so that there exists an incredible opportunity for the future discovery of new glass formulations with unusual functionalities.

However, although it offers a large room for improvement, the astronomical number of possible glass compositions is also a challenge. Indeed, such a large parametric space renders traditional Edisonian discovery approach based on trial-and-error largely inefficient. To accelerate the discovery of new glass formulations, it is necessary to decode the *Glass Genome*, that is, to decipher how the properties of glasses are controlled by their underlying composition and structural features (the glass "genes").

### *1.3 Topological constraint theory (TCT)*

Over the past decades, ***topological constraint theory*** (TCT) [10] has been an invaluable tool to help to investigate these problems and gain a better understanding of the linkages between structure and properties in disordered materials. The success of TCT is based on the fact that many macroscopic properties depend primarily on the atomic network topology of materials. In turn, simplifying a complex material into simpler networks of nodes (atoms) that are interconnected to each other via some constraints (chemical bonds) makes it possible to develop models that can be used to analytically predict material properties. Among others, TCT has been used to predict the properties of glasses [11], elucidate the origin of concrete creep [12], identify durable phase-change semiconducting materials [13], understand the effect of irradiation on minerals [14], and examine the origin of protein folding [15]. One of the most popular examples of the power of TCT is offered by Corning® Gorilla® Glass (a scratch- and damage-resistant glass used in more than 5 billion smartphones and tablets), which was designed *in silico* through the optimization of its atomic network topology before anything was actually melted in a laboratory [16,17].

In the following, we provide a general introduction to glass science and topological constraint theory, and review how TCT can be used to predict the properties of glasses to pinpoint promising compositions featuring optimal functionalities.

## 2. Introduction to glass science
### *2.1 The glassy state*

Despite their critical importance in various applications, the nature of glasses—not truly a solid, neither a liquid—remains poorly understood [18]. The *V–T* diagram shown in **Fig. 2** is probably the most instructive plot in glass science [19]. This plot depicts the evolution of volume (or, equivalently, enthalpy) as a function of temperature for glass-forming systems. At high temperature, materials tend to exist in the form of a *liquid* (or eventually gas) state. In this state, due to thermal expansion, liquids typically significantly expand with increasing temperature. At this point, the system is at stable equilibrium. If cooled slowly enough, liquids can undergo crystallization at convert into *crystals*. This transition is sharp (first-order transition), occurs at fixed temperature (melting temperature $T_m$), and is associated with a discontinuity (decrease) in volume (and enthalpy, see **Fig. 2**). Crystals are also at stable equilibrium and typically represent the most compact and stable state of matter. Crystallization occurs in two stages, namely, nucleation (i.e., the formation of crystal nuclei) and crystal growth. However, since crystallization is not instantaneous (i.e., it requires an energy barrier to be overcome), liquids can be cooled below $T_m$ while not crystallizing. At this point, they are referred to as *supercooled liquids*. Supercooled liquids are in metastable equilibrium, that is, they occupy a local minimum position in the enthalpy landscape. At this point, there exists a thermodynamic driving force toward crystallization since the free energy of the crystal is lower than that of the supercooled liquid. However, if cooling continues, the viscosity of the melt tends to increase exponentially with decreasing temperature [20]. At some point, the viscosity becomes so high that relaxation becomes kinetically impossible and the melt starts to behave like a solid. A *glass* is formed. This transition is continuous and manifests itself by a gradual decrease in the slope of $V(T)$ (i.e., thermal expansion coefficient), which eventually becomes comparable to that of a crystal. The temperature around which the gradual break of slope occurs is referred to as the fictive temperature $T_f$, as, to the first order, a glass can be considered as a frozen supercooled liquid cooled down to $T_f$. In the glassy state, the system is out-of-equilibrium, that is, it continually wants to relax toward the metastable supercooled liquid state, but relaxation largely exceeds the observation time (i.e., it can greatly exceed the age of the universe) [21]. Based on this, glass has been defined as "*a nonequilibrium, non-crystalline state of matter that appears solid on a short time scale but continuously relaxes towards the liquid state*" [18]. This definition distinguishes glasses (which relax toward the supercooled liquid state upon heating) from amorphous solids (which tend to crystallize upon heating) [22].

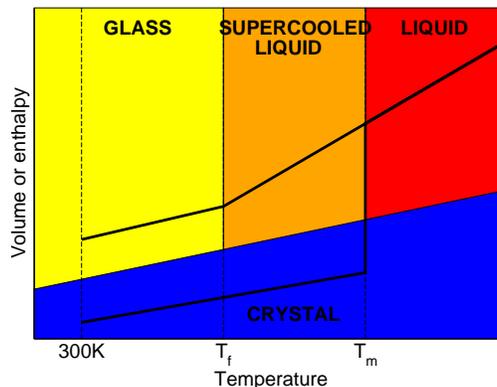

**Fig. 2:** Schematic showing the evolution of volume (or enthalpy) as a function of temperature in a glass-forming system. $T_m$ and $T_f$ are the melting and fictive temperature, respectively.

### *2.2 Network formers vs. network modifiers*

A fundamental question in glass science is to understand what makes it possible for a substance to avoid crystallization upon cooling below its melting point, that is, to understand the origin of glass-forming ability. For oxide glasses, some useful insights can be gained by considering the chemical composition of the system. The cations A that form the network of oxide glasses can be classified based on the energy of the A–O bonds. Namely, the cations A are classified as network formers and network modifiers if the A–O bond energy is higher than 330 kJ/mol and lower than 250 kJ/mol, respectively [19]. Other cations are classified as intermediate. Network-forming species comprise Si, B, Ge, Al, P, etc., whereas network-modifying species comprise Ca, Mg, Li, Na, K, etc. This classification is based on the idea that, to avoid crystallization during cooling, supercooled liquids must have a viscosity that is high enough to prevent the atoms from easily reorganizing. This requires the existence of strong interatomic bonds. As such, network formers form the backbone of glasses, whereas network modifiers tend to depolymerize them.

### *2.3 Zachariasen rules of glass formation*
The paper "The Atomic Arrangement in Glass" published by Zachariasen in 1932 is largely considered as the birth of modern glass science [23]. In this contribution, Zachariasen introduced a series of rules dictating the ability of a given compound to form a glass. Zachariasen's approach is based on several observations and ideas. First, the properties (stiffness, density) of glasses are fairly similar to those of their isochemical crystals. This suggests that the atomic structure of glasses cannot be too different from that of crystals. Further, although crystals exhibit a lower free energy than glasses, this difference of free energy cannot be too large—otherwise, there would exist a very large driving force promoting crystallization. For all these reasons, Zachariasen proposed that the local atomic structure of glasses must be fairly similar to that of crystals. Second, to avoid crystallization, glasses must macroscopically rigid. To this end, they must form an extended three-dimensional atomic network, that is, the atomic connectivity must be high enough. Third, glasses do not exhibit any sharp peak upon X-ray diffraction. As such, the atomic network of glasses must be random at distance larger than a few atomic bond distances. This requires an open structure that is flexible enough to yield some long-range randomness. An atomic connectivity that would be too high would not allow such flexibility.

Based on these ideas, Zachariasen proposed a series of four rules to describe the ability of a given oxide compound $A_mO_n$ to form a glass [23]:
(1) The oxygen atoms O must be linked to no more than two cations A.
(2) The number of O neighbors around A cations (i.e., their coordination number) must be small, between 3 and 4.
(3) The cation polyhedra must share corners with each other, not edges or faces.
(4) At least three corners of the cation polyhedra must be shared.

Rules (1), (2), and (3) ensure that the atomic network of glasses is flexible enough to enable the formation of a random (i.e., non-periodic) network. In turn, rule (4) ensures that the atomic connectivity is high enough to form a continuous three-dimensional network that is rigid enough to resist crystallization. Topological constraint theory also relies on this idea that glass must exhibit an atomic connectivity that is neither too low, neither too high to be able to avoid crystallization.

### 3. Topological constraint theory and glass rigidity
### *3.1 Stability of mechanical trusses*
The topological nanoengineering of glasses takes its root in the study of the stability of mechanical trusses, as initially developed by Maxwell and Lagrange [24], as described in the following. Let us first consider the three simple trusses presented in **Fig. 3**. The left one is clearly flexible as it can be freely deformed with no external energy, whereas the other two trusses are rigid (note that, in this context, rigid means "not flexible" and does not imply that the solid has an infinite stiffness). In general, the degree of flexibility of a given truss can be determined by comparing the number of mechanical constraints $N_c$ (i.e., the number of red sticks in this case) and the initial number of degrees of freedom of the truss nodes. In two-dimension, the initial number of degrees of freedom is given by $2N$ (i.e., two translation directions), where $N$ is the number of nodes. The number of remaining internal modes of deformation (or floppy modes) present within the truss $F$ after the application of the constraints is then given by:

$$F = 2N - N_c - 3 \quad \text{(Eq. 1)}$$

which arises from the fact that, starting from the situation wherein each node can freely move along two directions, each constraint $N_c$ removes one internal degree of freedom. Note that the term "3" corresponds to the three macroscopic degrees of freedom of a rigid structure in two-dimension (i.e., two translations and one in-plane rotation). Based on this, one finds $F = 1$, 0, and –1 for the three trusses presented in **Fig. 3** (see **Tab. 1**). This denotes that the left truss exhibits one internal mode of deformation, whereas the middle truss cannot be deformed. In contrast, the value $F = –1$ for the right truss denotes that, in this structure, there are more constraints than degrees of freedom. In that situation, all the constraints cannot be satisfied at the same time—just like the angles of a triangle cannot be varied to arbitrary values if the dimensions of the three edges are already fixed. In this case, some internal eigenstress will be observed as some of the constraints will be under tension whereas others will be under compression. Note that such internal eigenstress does not result in any macroscopic stress as the constraints under tension and compression mutually compensate each other—so that the overall structure is at zero macroscopic stress (see **Fig. 4**).

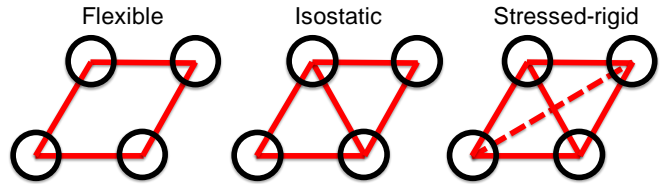

**Fig. 3:** The three states of rigidity of a mechanical truss. The dashed red line denotes a redundant constraint that is here under tension.

**Tab. 1:** Description of the stability of the three trusses presented in **Fig. 3**.

| Truss considered | Flexible (left) | Isostatic (middle) | Stressed-rigid (right) |
|---|---|---|---|
| **Initial # of degrees of freedom** | 8 | 8 | 8 |
| **# of constraints** | 4 | 5 | 6 |
| **# of internal modes of deformation** | 1 | 0 | 0 |
| **# of mutually-dependent constraints** | 0 | 0 | 1 |

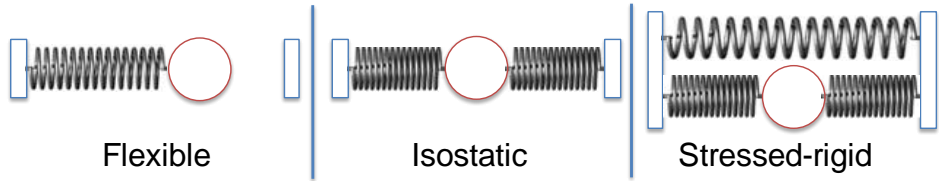

**Fig. 4:** Illustration of the origin of the internal eigenstress that is present in stressed-rigid structures.

In a similar fashion, the number of internal modes of deformation of a three-dimensional truss is given by:

$$F = 3N - N_c - 6 \quad \text{(Eq. 2)}$$

where $3N$ is the initial number of degrees of freedom of the nodes (i.e., before the application of the constraints) and the term "6" corresponds to the 6 macroscopic degrees of freedom of a rigid structure (i.e., three translations and three rotations). In general, mechanical trusses can be classified as (i) flexible if $F > 0$, (ii) stressed-rigid if $F < 0$, and (iii) isostatic (or statically-determinate) if $F = 0$, wherein flexible trusses exhibits some internal modes of deformation, stressed-rigid trusses exhibit some internal eigenstress, and isostatic trusses are rigid but free of stress.

### 3.2 Application to atomic networks

These concepts of structural stability were used by Phillips to establish topological constraint theory in 1979 [10] and refined by Thorpe in 1983 [25]. The main idea is that molecular networks can be seen as mechanical trusses, wherein the atoms are the nodes and the chemical bonds are the mechanical constraints (i.e., that prevent the relative motion between the atoms) [26]. In analogy with mechanical trusses, the chemical constraints effectively remove some of the initial degrees of freedom of the atoms (i.e., 3 per atom in three-dimensional networks). Molecular networks typically comprise two types of constraints: (i) the radial 2-body bond-stretching (BS) constraints that keep the inter-atomic distances fixed around their average values and (ii) the angular 3-body bond-bending (BB) constraints that keep the angles fixed around their average values (see **Fig. 5**). These bonds can be considered as little springs that prevent the relative motion between the atoms of the network [27]. In covalent networks wherein all the BS and BB constraints are intact (e.g., chalcogenide glasses), the number of BS and BB constraints created by each atom depend on its coordination number $r$. Namely, the number of BS constraints is given by $r/2$ (since each radial constraint is necessarily shared by 2 atoms) while the number of BB constraints is given by $2r - 3$ for $r \geq 2$ (since a triplet of atoms corresponds to 1 BB constraint and 2 new BB constraints are then needed to fix to the direction of each additional neighbor).

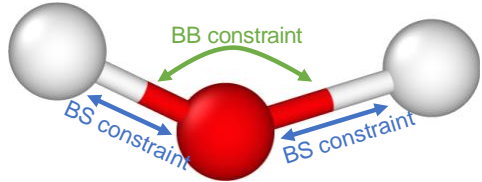

**Fig. 5:** Schematic illustrating the role of the radial bond-stretching (BS) and angular bond-bending (BB) constraints.

### 3.3 Mean field approximation

Let us consider a network of N atoms, where $N_i$ is the number of atoms having a coordination number $r_i \geq 2$. The total number of constraints $N_c$ is given by:

$$N_c = \sum_i \left[ N_i \left( \frac{r_i}{2} + 2r_i - 3 \right) \right] = \sum_i \left[ N_i \frac{5r_i}{2} \right] - 3N \quad \textbf{(Eq. 3)}$$

Due to high number of atoms in glass (typically on the order of $10^{23}$), it is more convenient to rely on a mean-field approximation and consider the average number of internal modes of deformation per atom $f = F/N$, the average number of constraints per atom $n_c = N_c/N$, and the fraction of each type of atom $x_i = N_i/N$. Following **Eq. 2**, the number of internal modes of deformation and constraints per atoms is then given by:

$$f = 3 - n_c - \frac{6}{N} = 3 - n_c \quad \textbf{(Eq. 4)}$$

$$n_c = \sum_i \left[ x_i \frac{5r_i}{2} \right] - 3 \quad \textbf{(Eq. 5)}$$

Note that the term 6/N becomes infinitely small for a large number of atoms and, thereby, can be ignored. Following Maxwell's stability criterion, glasses can then be classified as (i) **flexible** if $f > 0$ ($n_c < 3$), (ii) **stressed-rigid** if $f > 0$ ($n_c > 3$), and (iii) **isostatic** if $f = 0$ ($n_c = 3$). In analogy with mechanical trusses, flexible glasses are expected to exhibit some internal floppy modes of deformation (the number of floppy modes per atom being given by $f = 3 - n_c$). In contrast, stressed-rigid glasses are expected to present some internal stress due to the fact that some constraints mutually depend on each other (the number of redundant per atom being given by $n_c - 3$). In turn, isostatic glasses are free of both floppy modes and internal stress.

These relationships can be conveniently expressed in terms of the average coordination number $<r>$:

$$\langle r \rangle = \sum_i x_i r_i \quad \text{(Eq. 6)}$$

The number of constraints per atom can then be expressed as:

$$n_c = \frac{5\langle r \rangle}{2} - 3 \quad \text{(Eq. 7)}$$

The condition of isostaticity $n_c = 3$ is then satisfied for $<r> = 2.4$, i.e., the "magic" coordination number featured by isostatic networks (under the assumptions presented above). These results are summarized in **Fig. 6**.

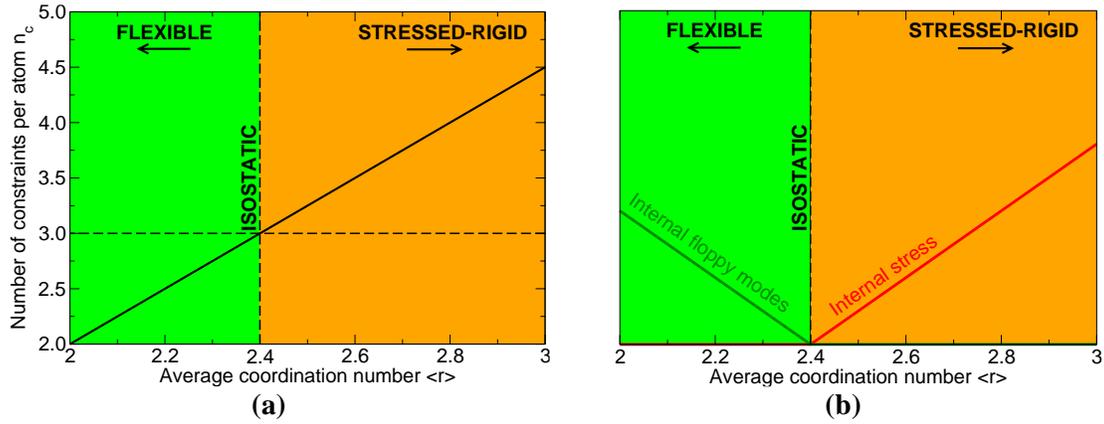

**Fig. 6:** **(a)** Schematic illustrating the evolution of the number of constraints per atom, **(b)** internal floppy modes, and internal stress as a function of the average coordination number in the glass.

### 3.4 Impact of 1-fold coordinated atoms
Note that **Eq. 7** is only valid if all BS and BB constraints are active and the coordination number of the atoms is always strictly larger than 1. In the presence of 1-fold coordinated atoms, the number of constraints per atom can be expressed as [28]:

$$n_c = \sum_{i \neq 1} \left[ x_i \left( \frac{r_i}{2} + 2r_i - 3 \right) \right] + \frac{x_1}{2} = \sum_i \left[ x_i \frac{5 r_i}{2} \right] + x_1 - 3 = \frac{5\langle r \rangle}{2} + x_1 - 3 \quad \text{(Eq. 8)}$$

where $x_1$ is the fraction of 1-fold coordinated atoms. Consequently, in the presence of 1-fold coordinated atoms, the isostatic condition $n_c = 3$ is achieved for:

$$\langle r \rangle = 2.4 - 0.4 x_1 \quad \text{(Eq. 9)}$$

Hence, the presence of 1-fold coordinated atoms results in a decrease in the average coordination number of isostatic glasses. This shift from the magic coordination number 2.4 is in agreement with experimental observations in amorphous hydrogenated silicon amorphous solids [28].

*3.5 Impact of temperature and pressure*

Strictly speaking, the previous constraint enumeration is only valid at zero temperature (i.e., low temperature) and zero pressure (i.e., ambient pressure). In turn, temperature and pressure can affect the topology of the atomic network of glasses, and, thereby, modify the number of constraints per atoms. To describe thermal effects, Mauro *et al.* introduced the idea of temperature-dependent constraints [29,30]. This is based on the idea that each constraint is associated with a given free energy and, therefore, can be active or thermally-broken based on the temperature. Namely, all constraints are active at low temperature and thermally-broken at infinite temperature (since the bonds can easily break or reform). In between, there exists an onset temperature $T_c$ at which a given constraint goes from being intact to thermally-broken. Note that different constraints exhibit different free energy and, therefore, can break at different temperatures. For instance, angular BB constraints are typically weaker than radial BS constraints and, hence, are associated to lower onset temperatures [31,32]. In a similar fashion, Bauchy and Micoulaut showed that, by altering the coordination of the atoms, pressure can increase or decrease the number of constraints per atom [33–35].

*3.6 Intermediate phase*

The previous equations rely on a mean-field approximation (see **Sec. 3.3**), which is based on the idea that the glass should be homogeneous—so that the average number of constraints is a representative metric. This yields a single isostatic threshold, that is, a single glass composition for which $n_c = 3$. However, this approach intrinsically cannot capture any local heterogeneity in the atomic topology [36–38]. Recently, Boolchand *et al.* suggested that, in many glass-forming systems, an isostatic state can be achieved for a continuous range of compositions rather than at a fixed threshold [39–42]. Boolchand's results suggest that, rather than a single flexible-to-stressed-rigid transition, glasses can experience two distinct transitions: (i) a flexible-to-rigid transition, that is, when the number of internal floppy modes of deformation becomes zero and (ii) an unstressed-to-stressed transition, that is, when an onset of eigenstress is observed within the structure. These two transitions define an ***intermediate phase***, wherein the atomic network is isostatic, that is, rigid (i.e., free of floppy modes) but unstressed. The existence of the intermediate phase has been attributed to some self-organization within the network, which reorganizes to become rigid while avoiding the onset of internal stress [33]. Entropy and weak Van der Waals interactions have also been suggested to play an important role [43]. Glasses belonging to the intermediate phase have been shown to exhibit unusual properties, e.g., an optimal space-filling tendency [42], a low propensity for relaxation [44], maximum resistance to creep [12], maximum fracture toughness [45], and maximum resistance to irradiation [46]. Note that the existence of the intermediate phase as well as that of potential structural signatures remains debated [14,47,48].

4. **Examples of constraints enumeration and application to glass-forming ability prediction**
   *4.1 Topological description of glass-forming ability*

In his pioneering contribution, Phillips predicted that isostatic glasses should exhibit optimal glass-forming ability [10], i.e., when the number of constraints exactly equals the number of atomic degrees of freedom. This can be explained by the fact that, in the flexible domain ($n_c < 3$), the atoms can easily reorganize toward lower energy states and, hence, glass can easily crystallize. In contrast, in the stressed-rigid domain ($n_c > 3$), the atomic network is locally unstable due to the existence of mutually-dependent constraints. In this regime, the redundant constraints exhibit some internal stress, that is, the weaker constraints yield to the stronger constraints by being under tension or compression. The existence of such internal stress acts as a driving force that stimulates relaxation toward lower energy states and, therefore, enhances the thermodynamic propensity for crystallization. This viewpoint is consistent with Zachariasen's description of glass formation, which also relies on the idea that the connectivity of the glass network should be high enough to form a continuous three-dimensional network, but low enough to remain flexible enough to form a random network (see **Sec. 2.3**). These ideas can also be expressed in terms of the underlying enthalpy landscape (see **Fig. 7**) [49–52]. Namely, in flexible glasses, the floppy modes of relaxation result in the formation of channels in between local minima within the enthalpy landscape, which facilitates atomic jumps from one state to another (see **Fig. 7a**). In contrast, in stressed-rigid glasses, the internal stress result in the existence of some local strain (elastic) energy that can promote the transition from one state to another [53–56] (see **Fig. 7c**). In turn, isostatic glasses are free of both floppy modes and internal stress and, hence, feature the highest kinetic and thermodynamic resistance to relaxation (and crystallization) [44].

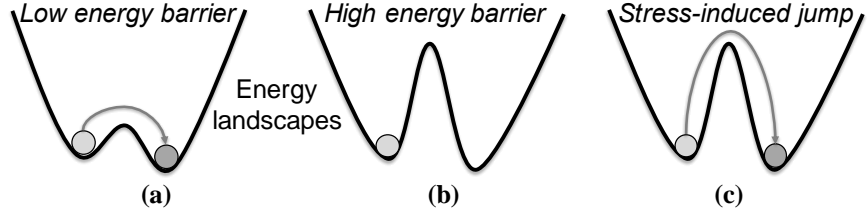

**Fig. 7:** Schematic illustrating the mechanism of relaxation toward a more stable state in the enthalpy landscape in **(a)** flexible, **(b)** isostatic, and **(c)** stressed-rigid glasses, respectively.

### *4.2 Chalcogenide glasses*

Historically, TCT was first applied to understand and predict the ability of chalcogenide alloys to form a glass when quenched fast enough [10]. In that sense, TCT offers a natural extension to the Zachariasen rules, which are limited to binary oxide systems [23]. Chalcogenide glasses (i.e., alloys of chalcogenide elements, e.g., Ge, Si, As, Sb, S, Se, Te, etc.) constitute an important class of glasses. In contrast to oxide glasses, chalcogenide glasses can form some homopolar bonds (e.g., Se–Se) and, hence, do not have to satisfy a fixed stoichiometry [57]. For instance, in contrast to the $Si_xO_{1-x}$ system (which can only exist for $x = 1/3$), $Ge_xSe_{1-x}$ glasses can be synthesized for continuous values of $x$ [40]. In many cases (although not always), the coordination number $r$ of the chalcogenide elements present in the glass is given by the 8–$N$ rule (or octet rule) [58]. For instance, one usually has $r_{Ge} = 4$, $r_{As} = 3$, and $r_{Se} = 2$ [59,60]. In the following, we focus on the glass-forming ability of the Ge–Se and As–Se systems.

#### *4.2.1 Ge–Se glasses*

We first focus on Ge–Se glasses. The structure of Ge–Se glasses comprises 4-fold coordinated tetrahedral Ge atoms and 2-fold coordinated Se atoms. Limited chemical order is observed as homopolar Ge–Ge and Se–Se bonds are observed [61]. **Tab. 2** details the constraints enumeration in $Ge_xSe_{1-x}$ glasses. As mentioned in **Sec. 3.2**, the number of radial BS and angular BB constraints can be obtained from the coordination number of each atom. The average number of constraints per atom $n_c$ is given by:

$$n_c = 7x + 2(1 - x) = 2 + 5x \quad \textbf{(Eq. 10)}$$

As expected, the number of constraints per atom increases with the fraction $x$ of highly coordinated Ge atoms. The isostatic composition can be identified by solving $n_c(x_{iso}) = 3$, which yields $x_{iso} = 20\%$. As such, $Ge_xSe_{1-x}$ is flexible for $x < 20\%$ and stressed-rigid for $x > 20\%$. This result can also be obtained by calculating the average coordination number:

$$\langle r \rangle = 4x + 2(1 - x) = 2 + 2x \quad \textbf{(Eq. 11)}$$

and solving the equation $<r>(x_{iso}) = 2.4$ (see **Sec. 3.3**), which also yields $x_{iso} = 20\%$ (i.e., a $GeSe_4$ glass).

**Tab. 2:** Constraints enumeration in Ge$_x$Se$_{1-x}$. The columns contain the types of atoms, their numbers (#), coordination number ($r$), number of bond-stretching (BS) constraints, number of bond-bending (BB) constraints, and total number of constraints per atom (BS+BB).

| Element | # | $r$ | # BS | # BB | # BS + BB |
|---|---|---|---|---|---|
| Ge | x | 4 | 2 | 5 | 7 |
| Se | 1 − x | 2 | 1 | 1 | 2 |

This result is a notable success for TCT as it provides an intuitive and elegant explanation to the experimental observation that Ge$_x$Se$_{1-x}$ exhibits maximum glass-forming ability around $x_{iso}$ = 20% (see **Fig. 8**). For instance, a GeSe$_4$ glass can be formed even by cooling a melt *in situ* in the furnace, whereas air- or water-quenching are required to form Se ($x$ = 0%) and GeSe$_2$ ($x$ = 33%) glasses [62]. These results are also in agreement with the fact that an intermediate phase wherein Ge–Se exhibits minimum non-reversible enthalpy at the glass transition is observed around $x$ = 20% [39] (see **Fig. 9a**). This suggests that, around the isostatic threshold, glasses exhibit maximum stability, i.e., minimum relaxation. This has been suggested to arise from the fact that (i) for $n_c < 3$ (flexible domain), relaxation is *facilitated* by the low-energy atomic modes of deformation, (ii) for $n_c > 3$, (stressed-rigid domain) relaxation is *stimulated* by the fact that the glass is unstable due to the presence of internal stress within the network, whereas (iii) for $n_c \approx 3$ (isostatic domain), the glass is both free of any internal modes of deformation and stress [63]. It is also interesting to note that Ge–Se glasses exhibit minimum molar volume around the isostatic threshold (see **Fig. 9b**). Such space-filling tendency has been observed for various glasses, which suggests that is a generic feature of isostatic glasses [42,64]. This can be explained as follows. Flexible Ge–Se glasses present some elongated chains of Se that reduce the local packing efficiency. In turn, stressed-rigid Ge–Se glasses exhibit a locked atomic network that cannot easily reorganize to efficiently fill the space.

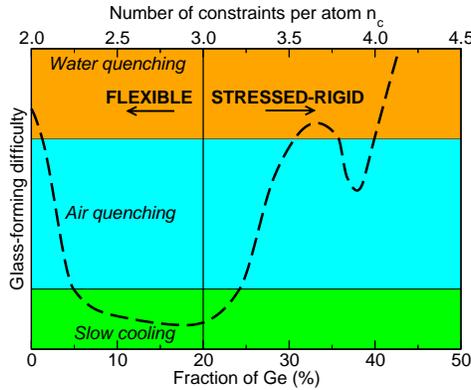

**Fig. 8:** Glass-forming difficulty of Ge–Se alloys as a function of composition and number of constraints per atom [62].

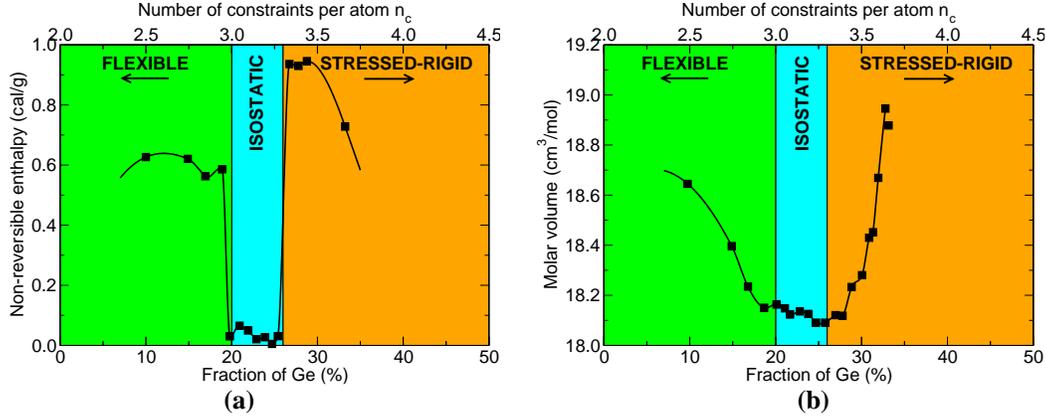

**Fig. 9: (a)** Non-reversible enthalpy at the glass transition (measured by modulated differential scanning calorimetry [65]) and **(b)** molar volume of Ge–Se glasses as a function of composition and number of constraints per atom [66].

*4.2.2 As–Se glasses*

We now focus on As–Se glasses. The structure of As–Se glasses comprises 3-fold coordinated pyramidal As atoms and 2-fold coordinated Se atoms [67,68]. **Tab. 3** details the constraints enumeration in $As_xSe_{1-x}$ glasses. The average number of constraints per atom $n_c$ is given by:

$$n_c = \frac{9}{2}x + 2(1-x) = 2 + \frac{5}{2}x \quad \textbf{(Eq. 12)}$$

As expected, the number of constraints per atom increases with the fraction $x$ of As atoms. The isostatic threshold is achieved at $n_c = 3$, which yields $x_{iso} = 40\%$ (i.e., a $As_2Se_3$ glass) [69]. Again, these predictions are supported by experimental results and, as in the case of Ge–Se, As–Se glasses present minimum non-reversible enthalpy and maximum space-filling tendency at the vicinity of the isostatic threshold [68].

**Tab. 3:** Constraints enumeration in $As_xSe_{1-x}$. The columns contain the types of atoms, their numbers (#), coordination number ($r$), number of bond-stretching (BS) constraints, number of bond-bending (BB) constraints, and total number of constraints per atom (BS+BB).

| Element | # | r | # BS | # BB | # BS + BB |
|---|---|---|---|---|---|
| As | x | 3 | 3/2 | 3 | 9/2 |
| Se | 1 – x | 2 | 1 | 1 | 2 |

*4.3 Network-forming oxides*

*4.3.1 SiO₂ glass*

We now focus on oxide glass-forming systems. We first discuss the case of glassy $SiO_2$—the archetypal structural basis for all silicate glasses. The structure of $SiO_2$ is made of 4-fold coordinated Si atoms that form some $SiO_4$ tetrahedra, which are connected to each other via their corners [70,71]. All the oxygen atoms at the corners of this unit are connected to two Si atoms and are referred to as bridging oxygen (BO) [72]. **Tab. 4** summarizes the constraints enumeration in $SiO_2$. As expected, the number of BS constraints created by each atom is half of their coordination number (see **Sec. 3.2**). The number of BB constraints created by Si atoms is 5, i.e., the number of independent O–Si–O angles that need to be fixed to define the tetrahedral environment. However, the inter-tetrahedral angle Si–O–Si has been noted to exhibit a broad distribution, which suggests that this angle is poorly constrained [73–77]. Consequently, no BB constraint is assigned to O atoms (or, in other words, this constraint is considered broken). Altogether, the number of constraints per atom $n_c$ is given by:

$$n_c = \frac{1 \times 7 + 2 \times 1}{1+2} = 3 \quad \textbf{(Eq. 13)}$$

which confirms that the excellent glass-forming ability of silica arises from the isostatic nature of its atomic network. Note that the broken nature of the Si–O–Si BB constraint is key in explaining the great glass-forming ability of silica [78]. Such thermal breakage of the BB constraints of O atoms is only observed in pure SiO$_2$ and was suggested to arise from the high glass transition temperature $T_g$ of silica (1200 °C)—so that this weak constraint is thermally broken at $T_g$ [31,79]. In contrast, this constraint is active is TeO$_2$, which explains its poor glass-forming ability [78]. Note that, in this case, despite the isostatic nature of SiO$_2$, the average coordination number is not equal to 2.4 as this relationship assumes that all BB constraints are intact (see **Sec. 3.2**).

**Tab. 4:** Constraints enumeration in SiO$_2$. The columns contain the types of atoms, their numbers (#), coordination number ($r$), number of bond-stretching (BS) constraints, number of bond-bending (BB) constraints, and total number of constraints per atom (BS+BB).

| Element | # | $r$ | # BS | # BB | # BS + BB |
|---------|---|-----|------|------|-----------|
| Si | 1 | 4 | 2 | 5 | 7 |
| O | 2 | 2 | 1 | 0 | 1 |

### 4.3.2 B$_2$O$_3$ glass

We now focus on B$_2$O$_3$, that is, the base structural unit of all borate glasses. The structure of B$_2$O$_3$ is made of 3-fold coordinated B atoms that form some trigonal BO$_3$ units, which are connected to each other via their corners [19]. **Tab. 5** summarizes the constraints enumeration in B$_2$O$_3$. In contrast to the case of SiO$_2$, the BB constraint associated to the B–O–B angle is here considered intact on account of the low glass transition temperature of this glass. The number of constraints per atom $n_c$ is given by:

$$n_c = \frac{2 \times 9/2 + 3 \times 2}{2 + 3} = 3 \qquad \textbf{(Eq. 14)}$$

which establishes the isostatic nature of the atomic network of B$_2$O$_3$ and explains its excellent glass-forming ability. This is in agreement with the fact that the average coordination number of B$_2$O$_3$ is 2.4.

**Tab. 5:** Constraints enumeration in B$_2$O$_3$. The columns contain the types of atoms, their numbers (#), coordination number ($r$), number of bond-stretching (BS) constraints, number of bond-bending (BB) constraints, and total number of constraints per atom (BS+BB).

| Element | # | $r$ | # BS | # BB | # BS + BB |
|---------|---|-----|------|------|-----------|
| B | 2 | 3 | 3/2 | 3 | 9/2 |
| O | 3 | 2 | 1 | 1 | 2 |

### 4.3.3 P$_2$O$_5$ glass

Finally, we focus on P$_2$O$_5$, which is the base structural unit for all phosphate glasses. The structure of P$_2$O$_5$ is made of 4-fold tetrahedral PO$_4$ units [80]. However, in contrast to silica, only three of the O corners of the tetrahedra are shared and one O atom is terminating (i.e., non-bridging oxygen, or NBO) [80]. As such, there is 1 NBO and 3/2 BO per P atom, respectively. **Tab. 6** summarizes the constraints enumeration in P$_2$O$_5$. Note that, in this case, since the two types of O atoms (BO and NBO) feature a different topology, they must be treated in a distinct fashion in the enumeration (i.e., as if they were different elements). The number of constraints per atom $n_c$ is given by:

$$n_c = \frac{2 \times 7 + 3 \times 2 + 2 \times 1/2}{2 + 5} = 3 \qquad \textbf{(Eq. 15)}$$

which, again, demonstrates the isostatic nature of the atomic network of P$_2$O$_5$ and explains its good glass-forming ability. Note that, despite the isostatic nature of P$_2$O$_5$, the average coordination number is not equal to 2.4 as this relationship assumes that each atom has a coordination number that is equal or larger than 2 (see **Sec. 3.2**).

**Tab. 6:** Constraints enumeration in $P_2O_5$. The columns contain the types of atoms, their numbers (#), coordination number ($r$), number of bond-stretching (BS) constraints, number of bond-bending (BB) constraints, and total number of constraints per atom (BS+BB).

| Element | # | r | # BS | # BB | # BS + BB |
|---|---|---|---|---|---|
| P | 2 | 4 | 2 | 5 | 7 |
| O | 5 | - | - | - | - |
| BO | 3 | 2 | 1 | 1 | 2 |
| NBO | 2 | 1 | 1/2 | 0 | 1/2 |

*4.4 Impact of network modifiers and the example of sodium silicate glass*

In contrast to the network-forming species, network-modifying atoms tend to depolymerize the network by creating some weak bonds [81]. In the following, we now review how TCT can describe the glass-forming ability of a silicate glass comprising some network modifiers by taking the example of sodium silicate, an archetypical model for all alkali silicate glasses [82]. Such glasses are technologically important as they can be strengthened using ion exchange (e.g., Corning® Gorilla® glass) [63,83–85]. Starting from the base topology of glassy silica, Na atoms tend to depolymerize the silicate network by creating some NBO atoms (1 NBO per Na atom) [19]. **Tab. 7** summarizes the constraints enumeration in the alloy $(Na_2O)_x(SiO_2)_{1-x}$. As in the case of $P_2O_5$, it is here necessary to distinctly account for the BO and NBO atoms as they exhibit a different topology. Note that, here, the number of BS constraints created by Na differs from their coordination number (which is found to be around 6 in sodium silicate glasses) [86–88]. Indeed, in this case, atomistic simulations suggested that, despite having a coordination of around 6, Na atoms are preferentially bounded to the nearest NBO atom and, hence, exhibit only 1 BS constraint [32]. This illustrates the fact that the number of BS constraints created by an atom is not always given by the geometrical coordination number, which makes it necessary to have an accurate knowledge of the glass structure to meaningfully enumerate the constraints. Also note that even a small fraction $x$ of $Na_2O$ results in a drop of the glass transition temperature of the glass with respect to that of pure silica [89]. Consequently, in contrast to silica, the BB constraint associated to BO is not considered as being thermally-broken anymore [32]. Finally, note that, due to the ionic and non-directional nature of the Na–O bonds, the Si–NBO–Na angle is very poorly defined [86]. Hence, no BB constraint is assigned to this angle. Altogether, the number of constraints per atom $n_c$ is given by:

$$n_c = \frac{(1-x)\times 7 + 2x\times 1/2 + (2-3x)\times 2 + 2x\times 1}{(1-x)+(2-x)+2x} = \frac{11-10x}{3} \quad \textbf{(Eq. 16)}$$

As expected, the rigidity (i.e., $n_c$) of the glass decreases upon the addition of Na, that is, as the network becomes more and more depolymerized. An isostatic atomic network is then obtained for $n_c = 3$, which is achieved at $x_{iso} = 20\%$. As such, sodium silicate is stressed-rigid ($n_c > 3$) for $x < 20\%$ and flexible ($n_c < 3$) for $x > 20\%$. These predictions are in agreement with experimental results, as sodium silicate exhibits a minimum in enthalpy relaxation around $x_{iso} = 20\%$ (see **Fig. 10**) [89,90]. Sodium silicate glasses also exhibit a sharp increase in elastic energy (calculated from the variation in density upon annealing) for $x > 20\%$, which denotes an onset of internal flexibility, in agreement with the present predictions [89,90]. These results are also in agreement with the fact sodium silicate glasses tend to show some degree of phase separation (i.e., low glass-forming ability) for $x < 20\%$ [91], which may arise from the presence of internal stress within the structure. Altogether, this analysis illustrates how TCT can be used to describe the rigidity of modified silicate glasses.

**Tab. 7:** Constraints enumeration in $(Na_2O)_x(SiO_2)_{1-x}$. The columns contain the types of atoms, their numbers (#), coordination number ($r$), number of bond-stretching (BS) constraints, number of bond-bending (BB) constraints, and total number of constraints per atom (BS+BB).

| **Element** | **#** | **$r$** | **# BS** | **# BB** | **# BS + BB** |
|---|---|---|---|---|---|
| Si | $1-x$ | 4 | 2 | 5 | 7 |
| O | $2-x$ | - | - | - | - |
| Na | $2x$ | 1 | 1/2 | 0 | 1/2 |
| BO | $2-3x$ | 2 | 1 | 1 | 2 |
| NBO | $2x$ | 1 | 1 | 0 | 1 |

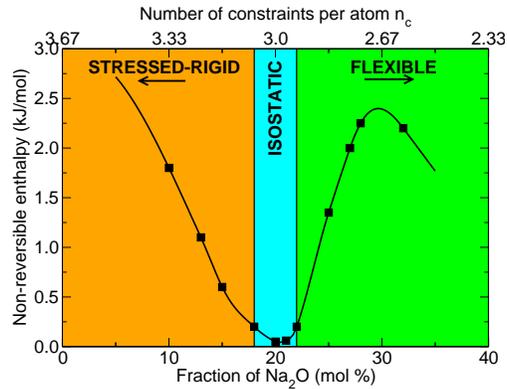

**Fig. 10:** Non-reversible heat flow at the glass transition (measured by modulated differential scanning calorimetry) in $(Na_2O)_x(SiO_2)_{1-x}$ glasses as a function of composition and number of constraints per atom [89].

### *4.5 Interest of molecular dynamics simulations*

Although the number of constraints per atom can be calculated analytically for select glasses, it requires an accurate knowledge of the glass structure that is not always available. In addition, several issues can render the constraints enumeration challenging.

(1) The coordination number of each species is not always known. For instance, in the case of borosilicate glasses, B atoms can be 3-fold or 4-fold coordinated depending on composition [92,93] and Ca atoms typically exhibit a coordination that is between 4 and 8, thereby not following the Octet rule [94]. Moreover, it has been shown that the effective number of BS constraints created by each atom is not always equal to their coordination number (i.e., as calculated by enumerating the number of neighbors inside the first coordination shell). In particular, Na atoms in sodium silicate glass have a coordination number around 6, but only show one active BS constraint [32].
(2) Isolated atoms or molecules (e.g., water molecules [95–97]) are not part of the atomic network and, hence, do not contribute to its rigidity. Therefore, they should not be taken into account in the constraint enumeration.
(3) Each constraint is associated with a given energy and can consequently be intact or broken depending on temperature, i.e., the amount of available thermal energy [29]. Hence, weaker angular constraints (like the Si–O–Na bond in sodium silicate) are broken even at 300 K [32,98]. For these reasons, one cannot just rely on unproven guesses to enumerate the number of constraints.

Molecular dynamics simulations, which offer a full access to the structure and dynamics of the atoms, provide a valuable tool to tackle these difficulties. To this end, a general enumeration method has been developed, which allows one to compute the number of constraints per atom in network glasses. This method has been widely applied to chalcogenide and oxide glasses [27,32,34,69,73,99] and, more recently, to atomic-scale models of cement hydrates [45,96] This enumeration method is based on the analysis of atomic trajectories obtained through molecular dynamics simulations. Since the nature of the constraints imposed on the atomic motion is not known *a priori*, the opposite approach is adopted, that is, by looking at the motion of each atom and deducing the underlying constraints that cause this motion. In other words, an active constraint would maintain bond lengths or angles fixed around their average values, whereas a large atomic motion implies the absence of any underlying constraint (see **Fig. 11**). Specifically, to assess the number of BS constraints that apply to a central atom, the radial excursion of each neighbor is computed. A small radial excursion implies the existence of an underlying constraint that maintains the bond length fixed around its average value. On the contrary, a large radial excursion implies a broken constraint. The number of BB constraints can be accessed in the same fashion, that is, by analyzing the angular excursion of each neighbor. The detailed implementation of this method is reported in [32,96,100] and unambiguously discriminates intact from broken constraints, including those created by Na and Ca atoms. As such, molecular dynamics can be used to inform topological constraint theory so that the constraints enumeration is based on an accurate structural base rather than simple guesses.

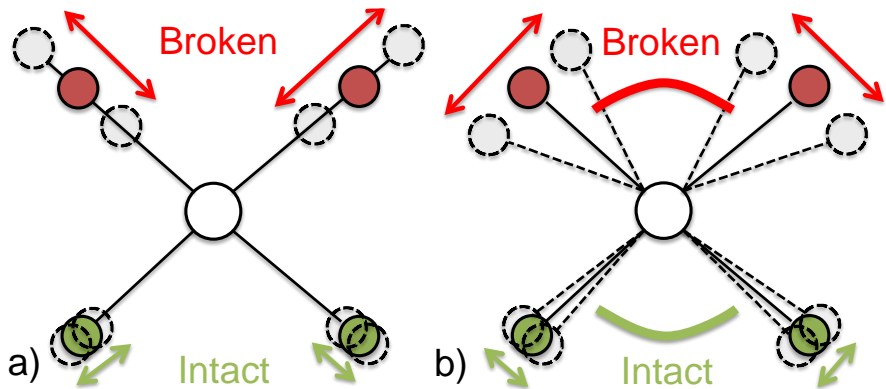

**Fig. 11:** Illustration of the usage of molecular dynamics to compute the number of (a) radial BS and (b) angular BB constraints per atom [32,96].

## 5. Applications of topological constraint theory for the prediction of glass properties
### 5.1 General principles
Besides glass-forming ability, topological constraint theory has been extensively applied toward the prediction of the compositional dependence of glass properties. These topological models are based on the following approaches:
  (1) By capturing the important connectivity of glasses while filtering out less relevant structural details, topological constraint theory can be used to capture the first-order contribution of the structure of the atomic network on macroscopic properties. This makes it possible to simplify complex disordered network into more simple mechanical trusses, thereby allowing the development of analytical predictive models. These models rely on (i) the knowledge of the glass structure as a function of composition, so that one can analytically calculate the number of constraints per atom $n_c(x)$ as a function of the composition $x$ of the glass, and (ii) a model relating $n_c$ to a given macroscopic property $P(n_c)$. The combination of (i) and (ii) yields an analytical model $P(x)$. Hence, the number of constraints per atom acts as a reduced-order parameter to facilitate the understanding of the linkages between structure and macroscopic properties. For instance, this type of models is used for predicting glass hardness as a function of composition (see **Sec. 5.2**).
  (2) For select properties, no direct analytical model relating $n_c$ and $P$ are yet available. However, $P$ is sometimes known to be minimized or maximized for optimally-constrained isostatic glasses, e.g., $P_{max} = P(n_c = 3)$. As such, this type of model cannot predict the compositional dependence of a property but can be used to pinpoint promising compositions exhibiting optimal properties. These models rely on (i) the knowledge of the glass structure as a function of composition so that one can analytically calculate the number of constraints per atom $n_c(x)$ as a function of the composition $x$ of the glass and (ii) the ability to solve $n_c(x_{iso}) = 3$. To predict optimal composition(s) $\{x_{iso}\}$. For instance, this type of models is used for predicting glass compositions exhibiting maximum fracture toughness (see **Sec. 5.3**).

In the following, we briefly review a selection of the topological models that are available in the literature.

### 5.2 Hardness

Hardness characterizes the resistance of materials to permanent deformations [101,102]. Smedskjaer, Mauro, and Yue proposed that the hardness of glasses can be predicted from the knowledge of the number of constraints per atom [11]. They introduced the following formula:

$$H = \left(\frac{\partial H}{\partial n_c}\right)[n_c - n_{crit}] \quad \textbf{(Eq. 17)}$$

This formula is based on the idea that (i) a glass needs a critical minimum number of constraints $n_{crit}$ to be cohesive and that (ii) each additional constraints should contribute to increase hardness. The value $n_{crit} = 2.5$ (i.e., the minimum number of constraints that are needed to achieve rigidity in two dimensions) was found to be appropriate [11], whereas the $\partial H/\partial n_c$ term is a fitting parameter that depends on the indenter geometry, the indentation load, and the glass family being considered (but not the specific glass composition). This model has been found to yield accurate hardness predictions for various oxide glasses [11,103]. Recently, it was suggested that BS and BB constraints may not contribute in the same fashion to hardness. Specially, the hardness of various materials has been found to preferentially scale with the number of angular BB constraints per atom rather than the total number of constraints per atom (see **Fig. 12**) [31]. This has been attributed to the fact that, upon indentation loading, the atomic network deforms by following the lowest energy path, that is, by breaking the weaker angular BB constraints rather than the stronger radial BS constraints. It has also recently been suggested that hardness should depend on the number of constraints per unit of volume rather than the number of constraints per atom [104].

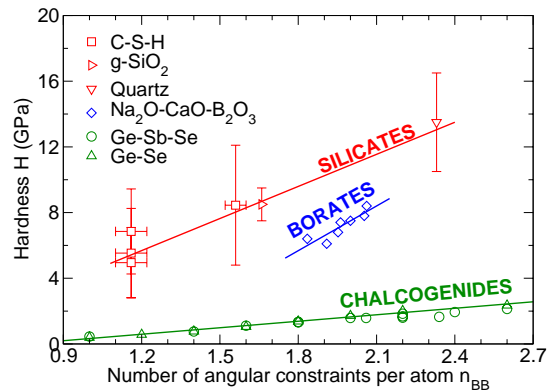

**Fig. 12:** Hardness of various oxide and chalcogenide phases as a function of their number of angular BB constraints per atom [11,31,96,105–115].

### 5.3 Fracture toughness

Fracture toughness characterizes the resistance of materials to fracture [116–122]. Although no topological model predicting the compositional dependence of the fracture of glasses has been proposed thus far, various glasses have been noted to exhibit maximum fracture toughness at the isostatic threshold (see **Fig. 13**) [45,108]. This has been suggested to arise from the fact that (i) flexible glasses ($n_c < 3$) exhibit low cohesion (low surface energy) due to their low connectivity, (ii) stressed-rigid glasses ($n_c > 3$) break in a brittle fashion as their high connectivity prevents any ductile atomic reorganization, whereas (iii) isostatic glasses ($n_c = 3$) exhibit the best balance between cohesion and ability to plastically deform [45]. Specifically, isostatic glasses have been noted to exhibit a maximum propensity for crack blunting, which contributes to postponing fracture [45].

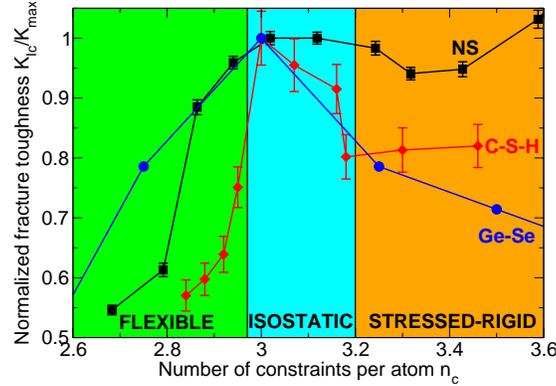

**Fig. 13:** Fracture toughness (normalized by the fracture toughness at $n_c = 3$) of densified sodium silicate glasses (NS), calcium–silicate–hydrates (C–S–H), and Ge–Se glasses as a function of their number of constraints per atom [45,116,122,123].

### 5.4 Viscosity, fragility, and glass transition temperature

Predicting the viscosity of glass-forming melts is critical for various applications [20,81]. However, such predictions are complex as (i) the viscosity varies by many orders of magnitude with temperature and (ii) it is sensitive to small changes in composition or structure [81]. Most mathematical models used to describe the viscosity of supercooled liquids are empirical and based on the Vogel-Fulcher-Tamman (VFT) equation:

$$\eta(T) = A \exp\left[\frac{B}{R(T-T_0)}\right] \quad \textbf{(Eq. 18)}$$

where $R$ is the perfect gas constant, $T$ is the temperature, and $A$ and $T_0$ some empirical parameters, which depend on the composition of the glass. Although the VFT equation has met notable success, it gives an incorrect description of the asymptotic Arrhenius behavior of viscosity at low temperature. To overcome this limitation, they MYEGA equation [20] was recently proposed to describe the temperature dependence of liquid viscosity:

$$\log \eta(T) = \log \eta_\infty + (12 - \log \eta_\infty)\frac{T_g}{T} \exp\left[\left(\frac{m}{12-\log \eta_\infty} - 1\right)\left(\frac{T_g}{T} - 1\right)\right] \quad \textbf{(Eq. 19)}$$

where $T_g$ is the glass transition temperature, defined by $\log \eta(T_g) = 10^{12}$ Pa•s, $\eta_\infty$ is the extrapolated infinite-temperature limit of liquid viscosity (universally found to be equal to $10^{-2.9}$ Pa•s), and $m$ is the liquid fragility index [124,125]:

$$m = \left.\frac{d \log \eta}{d(T_g/T)}\right| T = T_g \quad \textbf{(Eq. 20)}$$

As opposed to the VFT equation, this model has a clear physical foundation based on the temperature-dependence of the configurational entropy (given by the Adam-Gibbs equation) and offers more accurate predictions at low temperature. In addition, it only relies on two physical parameters, namely, the fragility index and the glass transition temperature. As such, the composition-dependence of the viscosity is entirely captured by that of $T_g$ and $m$. Analytical models have been developed to predict the composition dependence of the viscosity, $T_g$, and $m$ through the use of temperature-dependent topological constraint theory and the Adam-Gibbs equation [29,30,126,127]. Therefore, the prediction of the compositional dependence of these dynamical properties only relies on the mere knowledge of the number of topological constraints per atom $n_c$ in the network with respect to composition and temperature.

### 5.5 Dissolution kinetics

Predicting glass corrosion is important for various applications, including nuclear waste immobilization glasses, bioactive glasses, or supplementary cementitious materials [128–130]. It was recently proposed for the dissolution rate $K$ of glasses in dilute conditions (i.e., forward rate, far from saturation) is controlled by the topology of the atomic network [131] as follows:

$$K = K_0 \exp\left(-\frac{n_c E_0}{RT}\right) \quad \text{(Eq. 21)}$$

where $K_0$ is a rate constant that depends on the solution phase chemistry (i.e., the barrier-less dissolution rate of a completely depolymerized material for which $n_c = 0$) and $E_0$ = 20-to-25 kJ/mol is an energy barrier that needs to be overcome to break a unit atomic constraint. Based on this equation, the dissolution process is characterized by the effective activation energy:

$$E_A^{\text{eff}} = n_c E_0 \quad \text{(Eq. 22)}$$

The following atomistic picture was suggested to explain this model: starting from $n_c = 0$ (i.e., which would correspond to a fully depolymerized material), each new constraint per atom effectively reduce the dissolution kinetics by increasing the associated activation energy needed for bond rupture [131]. In details, it was proposed that $n_c$ serves as an indicator of steric effects acting in the atomic network, which prevent the reorganization and internal motion of the constituent species. Indeed, whether it occurs by hydrolysis or ion exchange, corrosion results in the formation of some local stress within the network. Namely, hydrolysis requires the formation of larger intermediate over-coordinated species (5-fold coordinated Si or three-fold coordinated O), whereas ion exchange requires some local opening of the network to enable the jump of mobile cations from one pocket to another [132]. In any case, the activation energy associated with these processes is controlled by the ability of the atomic network to locally reorganize to accommodate these local strains. In details, the resulting activation energy takes the form of the strain elastic energy that is applied by the rest of the network to resist the creation of this local defect [133]. This strain elastic energy is controlled by the local number of constraints per atom $n_c$, since each constraint acts as a little spring connecting the atoms. Therefore, $n_c$ characterizes the local stiffness of the atomic network, which tends to prevent the accommodation of local defects. This picture is in line with results from density functional theory, which have shown that the activation energy associated with the hydrolysis of inter-tetrahedra bridging oxygen atoms increases with the network connectivity and, therefore, rigidity [134]. As shown in **Fig. 14**, this model has since then been extensively validated over a broad range of silicate phases and is able to predict the dissolution kinetics of silicate phases over four orders of magnitude [12,135,130,136–141].

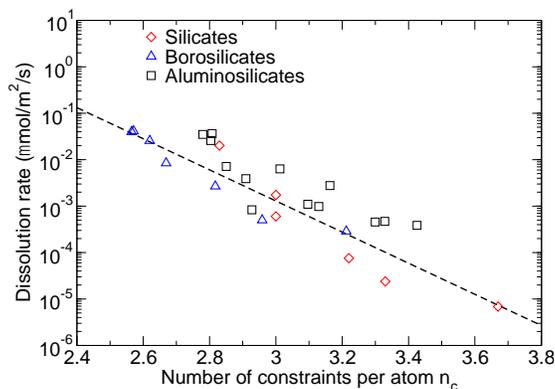

**Fig. 14:** Dissolution rate of various silicate phases as a function of the number of constraints per atom [130,136,139,141,131].

## *5.6 Other properties*

Several other properties have been shown to be correlated with the topology of the atomic network [142]. For instance, stiffness was found to scale with the number of constraints per atom in select systems [25,143]. Isostatic glasses were found to exhibit a stress-free character [144,145]. Isostatic glasses have also been found to exhibit optimal strengthening upon ion exchange [63,85,146,147]. Sub-critical crack growth was reported to be controlled by the atomic topology [148]. The degree of hydrophilicity/hydrophobicity of silica surfaces was found to be dictated by the topology of the glass surface [149]. Isostatic glasses were noted to exhibit maximum elastic volume recovery upon loading-unloading cycles [44,150]. Relaxation and aging were found to be minimized in isostatic glasses [33,35,44,151]. The propensity for creep was also found to be minimal in isostatic phases [12,44,152,153]. The thermal, mechanical, electrical, and optical properties of a-SiC:H thin films were demonstrated to be influenced by the topology of their atomic network [154]. Susceptibility was found to be maximal in isostatic granular systems [155,156]. The resistance to irradiation of silicate hydrates was found to be optimal in isostatic systems [14,46]. Fast-ion conduction was observed to offer a signature of the intermediate phase [157–159]. The performances of phase-change materials were noted to be controlled by their atomic topology [13]. Finally, protein folding was found to be controlled by the topology of their molecular architecture [15,160,161]. The wide variety of systems and properties for which topological constraint theory has been successfully applied denote its generic nature and suggests that the topological nanoengineering of materials is likely to yield exciting new developments in material science and engineering.

## 6. Conclusion

The various examples covered in the chapter highlight how topological constraint theory can be used as an effective tool to predict the properties of disordered materials or pinpoint optimized material compositions with tailored functionalities. It is especially well-suited for disordered systems with no fixed stoichiometry, for which traditional trial-and-error approaches are rendered inefficient due to the virtually infinite number of possible compositions. By capturing relevant structural features while filtering out the less relevant ones, topological approaches can largely simplify complex disordered structures and, thereby, facilitate the development of analytical predictive models—by capturing the essential structural features of atomic networks and using them as a reduced-order parameter to inform predictive models linking composition to properties. Topological models only rely on the accurate knowledge of the atomic structure of materials, which can be obtained by high-resolution experiments or atomistic simulations. As such, topological constraint theory exemplifies how a synergetic combination of experiments, simulations, and theoretical models can be used to decode the genome of glass, that is, by deciphering how glass' basic structural units control its macroscopic properties.

**Acknowledgements**

Matthieu Micoulaut, John C. Mauro, Morten M. Smedskjaer, Punit Boolchand, and Gaurav N. Sant are gratefully acknowledged for numerous discussions that form the basis of this Chapter. This work was supported by the National Science Foundation under Grants No. 1562066, 1762292, 1826420, and 1928538.